\documentclass[epj]{webofc}
\usepackage[utf8]{inputenc}
\usepackage[varg]{txfonts}   
\usepackage{booktabs}
\usepackage{xcolor}
\usepackage{braket}
\usepackage{bm}
\definecolor{darkred}{rgb}{0.4,0.0,0.0}
\definecolor{darkgreen}{rgb}{0.0,0.4,0.0}
\definecolor{darkblue}{rgb}{0.0,0.0,0.4}
\usepackage[bookmarks,linktocpage,colorlinks,
    linkcolor = darkred,
    urlcolor  = darkblue,
    citecolor = darkgreen]{hyperref}
\usepackage{subfigure}
\wocname{EPJ Web of Conferences}
\woctitle{Lattice2017}
\begin{document}
\newcommand{\beq}{\begin{eqnarray}}
\newcommand{\eeq}{\end{eqnarray}}
\newcommand{\nn}{ \nonumber}
\newcommand{\half}{\frac{1}{2}}
\newcommand{\chib}{{\bar \chi}}
\newcommand{\qb}{{\bar q}}
\newcommand{\Db}{{\bar {\cal D}}}
\newcommand{\D}{{\cal D}}
\newcommand{\ben}{\begin{enumerate}}
\newcommand{\een}{\end{enumerate}}
\newcommand{\bpsi}{{\bar \psi}}
\newcommand{\Bpsi}{{\bar \Psi}}
\newcommand{\etab}{{\bar \eta}}
\newcommand{\vev}[1]{{\langle #1 \rangle}}
\newcommand{\Dslash}{{\not \hspace{-4pt} D} }
\newcommand{\Dslashexp}{{\not \hspace{-1pt} D}}
\newcommand{\partialslash}{{\not \hspace{-4pt} \partial}}
\newcommand{\Aslash}{{\not \hspace{-4pt} A}}
\newcommand{\Pislash}{{\not \hspace{-4pt} \Pi}}
\newcommand{\Lcal}{{\cal L}}
\newcommand{\nnn}{ \nonumber \\ }
\newcommand{\ddd}{\nnn &&}
\newcommand{\angstrom}{\mbox{\normalfont\AA}}
\newcommand{\tr}{\mathop{{\hbox{Tr} \, }}\nolimits}
\newcommand{\Tr}{ {\rm Tr} \, }
\def\etal{{\it et al.}}
%
\selectlanguage{english}
\title{%
Direct detection of metal-insulator phase transitions using the
modified Backus-Gilbert method
}
\author{%
\firstname{Maksim} \lastname{Ulybyshev}\inst{1}\fnsep\thanks{Speaker, \email{maksim.ulybyshev@physik.uni-regensburg.de}}
\and
\firstname{Christopher} \lastname{Winterowd}\inst{2} \and
\firstname{Savvas}  \lastname{Zafeiropoulos}\inst{3,4,5}
}
\institute{%
Institut f\"ur Theoretische Physik, Universit\"at Regensburg, 93053 Regensburg, Germany
\and
University of Kent, School of Physical Sciences, Canterbury CT2 7NH, UK
\and
Thomas Jefferson National Accelerator Facility, Newport News, VA 23606, USA
\and
Department of Physics, College of William and Mary, Williamsburg, Virginia 23187-8795, USA
\and 
Institute for Theoretical Physics, Universit\"at Heidelberg, Philosophenweg 12, D-69120 Germany
}
\abstract{%
The detection of the (semi)metal-insulator phase transition can be extremely difficult if the local order parameter which characterizes the ordered phase is unknown. In some cases, it is even impossible to define a local order parameter: the most prominent example of such system is the spin liquid state. This state was proposed to exist in the Hubbard model on the hexagonal lattice in a region between the semimetal phase and the antiferromagnetic insulator phase. The existence of this phase has been the subject of a long debate.
In order to detect these exotic phases we must use alternative methods to those used for more familiar examples of spontaneous symmetry breaking. We have modified the Backus-Gilbert method of analytic continuation which was previously used in the calculation of the pion quasiparticle mass in lattice QCD. The modification of the method consists of the introduction of the Tikhonov regularization scheme which was used to treat the ill-conditioned kernel. 
This modified Backus-Gilbert method is applied to the Euclidean propagators in momentum space calculated using the hybrid Monte Carlo algorithm. In this way, it is possible to reconstruct the full dispersion relation and to estimate the mass gap, which is a direct signal of the transition to the insulating state. We demonstrate the utility of this method in our calculations for the Hubbard model on the hexagonal lattice. We also apply the method to the metal-insulator phase transition in the Hubbard-Coulomb model on the square lattice.
}
\maketitle
\section{Introduction}\label{intro}

Recently, methods commonly used in lattice QCD have been applied with great success to certain strongly-correlated electronic systems \cite{Drut:2009aj,Buividovich:2016qrg}. These applications have followed two distinct routes. For a certain class of systems, graphene in particular, a mapping to a low-energy effective field theory (EFT) allows the use of common lattice fermion discretizations which are employed to study the non-perturbative dynamics of the EFT \cite{Hands:2008id,Drut:2008rg,Buividovich:2012uk,DeTar:2016vhr,DeTar:2016dmj}. A second direction has been to directly simulate the many-body lattice Hamiltonian using path integral quantization and then applying the highly successful hybrid Monte Carlo approach to the theory \cite{Brower:2012zd,Buividovich:2012nx,Smith:2014tha}.  

Both of these approaches to strongly-correlated electronic systems have attempted to characterize their phase structure in a controlled, non-perturbative manner. In certain cases, such as the semimetal-insulator transition in the graphene EFT, or more precisely, the transition from the semimetal to the charge density wave (CDW) phase, the order parameter is known (the chiral condensate, $\vev{ \bpsi \psi}$). This allows one to identify the two phases by clearly established methods that are familiar to lattice practitioners. In other cases of physical interest, there is no known local order parameter. Two prominent examples are the Mott-Hubbard metal-insulator transition (see \cite{Imada:1998zz} for a comprehensive review) and the exotic and elusive spin-liquid phase \cite{PhysRevLett.95.036403,PhysRevB.76.035125,Meng2010,SciRep2012,Assaad:2013xua}. Progress in the understanding of both of the above-mentioned scenarios using lattice methods has been made by directly simulating the relevant lattice many-body Hamiltonians \cite{Ulybyshev:2017ped}. However, in order to do so, one must resort to looking at other, inherently non-local quantities, most notably spectral functions.

The long-standing problem of accessing spectral quantities through Euclidean correlators is notoriously difficult and subtle. Early attempts to attack this problem of analytic continuation (AC) used ans\"atze as well as Bayesian methods and had varied levels of success \cite{Karsch:1986cq,Asakawa:2000tr,Aarts:2007wj}. More recently, the Backus-Gilbert (BG) \cite{BackusGilbertOriginal} method has been employed in lattice QCD calculations \cite{Brandt:2015sxa} as well as in calculations for strongly-correlated electrons \cite{Boyda:2016emg,Ulybyshev:2017ped}. In this study, we apply a newly introduced modification of the BG method that employs the so-called Tikhonov regularization to a variety of lattice models: the Hubbard model on the hexagonal lattice and the Hubbard-Coulomb model on the square lattice. We do not find evidence in favor of the spin-liquid phase for the Hubbard model on the hexagonal lattice and we find that our new method for AC allows one to accurately determine the location and nature of the Mott-Hubbard transition.

\section{Methods}\label{sec-1}

 \subsection{Model}
 
We start by first defining the Hubbard-Coulomb Hamiltonian
\beq \label{SqHubbard}
\hat{\mathcal{H}} = - \kappa \sum_{\sigma} \sum_{\left \langle x, y \right \rangle} \left( \hat{c}^{\dagger}_{x,\sigma} \hat{c}_{y,\sigma} + \text{h.c.} \right) + \frac{1}{2}\sum_{x,y}  \hat{\rho}_x V_{x,y}\hat{\rho}_y,,
\eeq
where $\kappa$ is the hopping parameter and ${\left \langle x,y \right \rangle}$ denotes nearest neighbor sites. The creation and annihilation operators satisfy the following anticommutation relations
\beq \label{CommRelations}
\left\{ \hat{c}_{x,\sigma}, \hat{c}^{\dagger}_{y,\sigma'} \right\} = \delta_{x,y} \delta_{\sigma, \sigma'},
\eeq
where $x,y$ refer to the lattice site and $\sigma, \sigma'$ refer to the electron's spin. The two-body density-density interaction in (\ref{SqHubbard}) is completely general with $V_{x,y}$ a positive-definite matrix and $\hat{\rho}_x$ is the electric charge operator at site $x$. A pure on-site Hubbard model corresponds to the special case of a diagonal matrix $V_{x,y} = U \delta_{x,y}$. In our version of the Hubbard-Coulomb model, the two-body interaction is fully specified by the on-site coupling $U \equiv V_{0,0}$ and the nearest neighbor coupling $V \equiv V_{x,x+\delta}$, where $\bf{\delta}$ is a nearest neighbor vector, which determines the strength of the Coulomb tail \cite{Ulybyshev:2013swa,Ulybyshev:2017ped}.

The path integral formulation of the theory can be obtained by starting from the definition of the partition function and applying the Suzuki-Trotter decomposition
\beq
Z \equiv \Tr e^{-\beta\left( \hat{\mathcal{H}}\right)} = \Tr \left( e^{-\delta_{\tau} \hat{\mathcal{H}}_{0}} e^{-\delta_{\tau} \hat{\mathcal{H}}_{\text{int}}} e^{-\delta_{\tau} \hat{\mathcal{H}}_{0}} \dots \right) + O(\delta^2_{\tau}),
\eeq
where we have discretized the Euclidean time direction into $N_{\tau}$ slices such that $\delta_{\tau} \equiv \beta/N_{\tau}$ is the lattice spacing in the temporal direction. We also note that the above decomposition separates the kinetic term, $\hat{\mathcal{H}}_{0}$, from the interaction term, $\hat{\mathcal{H}}_{\text{int}}$. After further manipulations, including, for the four-Fermi term $\hat{\mathcal{H}}_{\text{int}}$, the introduction of a real scalar field $\phi$ via a Hubbard-Stratonovich transformation, we arrive at the following representation of the partition function
\beq \label{PathIntegral}
 Z = \int \mathcal{D}\phi  |\det M[\phi]|^2 e^{-S_B[\phi]},~S_B[\phi] = \tfrac{\delta_{\tau}}{2} \displaystyle \sum_{x,y,n}\phi_{x,n} V^{-1}_{x,y} \phi_{y,n}
\eeq
where $M$ is the inverse fermion propagator and $S_B[\phi]$ is the action of the bosonic Hubbard field. The fields carry both spatial indices as well as a temporal index which denotes the time-slice on which they reside. As per usual the bosonic fields satisfy periodic boundary conditions in the temporal direction $\phi_{x, n + N_{\tau}} = \phi_{x,n}$, while the fermionic fields satisfy anti-periodic boundary conditions $\psi_{x,n + N_{\tau}} = - \psi_{x,n}$ (which is encoded in the matrix $M$). For more details see \cite{Ulybyshev:2013swa,Ulybyshev:2017ped}.
The integrand in (\ref{PathIntegral}) is manifestly positive-definite due to the particle-hole symmetry of the model (\ref{SqHubbard}) at half-filling. This allows us to employ standard techniques of lattice gauge theory to sample systems with dynamical fermions in the absence of the sign problem. These algorithms broadly go under the name of hybrid Monte Carlo (HMC) \cite{Duane:1985hz,DeGrand:2006zz}.

Using our path integral formulation, the thermal expectation value of an observable $\hat{\mathcal{O}}$ is defined as 
\beq \label{Observable}
\vev{ \hat{\mathcal{O}}}_{\phi} = \frac{1}{Z} \int \mathcal{D}\phi \mathcal{O} |\det M[\phi]|^2 e^{-S_B[\phi]}.
\eeq
Information about the fermionic quasiparticles of the model in question can be obtained from the single-particle propagator
\beq \label{Propagator}
G(x,\tau) \equiv -\vev{ T c(x,\tau) c^{\dagger}(0,0) } = \vev{ M^{-1}_{x,\tau, 0,0} }_{\phi}, 
\eeq
where $T$ refers to time-ordering and we have suppressed spin indices and any position- or momentum-space labels. This theory has translational invariance in space and in Euclidean "time" as well as rotational symmetry. This implies that the propagators for up and down spins are the same and that they only depend on the spatial and temporal separation between the creation and annihilation operators appearing in (\ref{Propagator}).

 \subsection{Modified Backus-Gilbert Method}
 In order to obtain real-frequency information regarding a theory in thermal equilibrium, one must perform an analytic continuation of the correlator, such as (\ref{Propagator}), from discrete points on the imaginary axis (Matsubara frequencies) to the entire real line. This problem is ill-defined and it still remains a challenge to extract useful information from this procedure in a manner which requires the least amount of assumptions. The Green-Kubo (GK) relations express the connection between the Euclidean time correlator and the real-frequency spectral function. For this study, we write down the relevant GK relation for the fermionic quasiparticle
\beq \label{GreenKuboDOS}
   G(\tau) = \int_0^\infty d\omega K(\tau,\omega) A(\omega), \quad K(\tau,\omega) \equiv \frac{\cosh\left[\omega(\tau-\beta/2)\right]}{\cosh(\omega \beta/2)}.
 \eeq
In (\ref{GreenKuboDOS}), the spectral function $A(\omega)$ represents the density of states (DOS) for the fermionic excitations.
In order to invert the relation in (\ref{GreenKuboDOS}) using the BG method, we start by defining an estimator for the DOS
\beq \label{ConvolutionDeltaFunction}
     \bar{A}(\omega_0) = \int^{\infty}_0 d\omega \delta(\omega_0,\omega) A(\omega),
 \eeq
 which is the convolution of the exact DOS with the resolution function $\delta(\omega_0,\omega)$. Due to the linearity of the GK relations, one expresses the resolution function in the following basis
 \beq
   \delta(\omega_0,\omega) = \sum_{j} q_j(\omega_0) K(\tau_j, \omega), 
 \eeq
 which then, using (\ref{ConvolutionDeltaFunction}), implies
 \beq \label{Estimator}
   \bar{A}(\omega_0) = \sum_j q_j(\omega_0) G(\tau_j).
 \eeq
The coefficients $q_j(\omega_0)$ are determined by a minimization of the width of the resolution function around the frequency $\omega_0$
\beq
 D \equiv \int^{\infty}_0 d\omega (\omega-\omega_0)^2 \delta^2(\omega_0,\omega), \quad \int^{\infty}_0 d\omega \delta(\omega_0,\omega) = 1.
\eeq
The resulting expression is 
\beq
q_j(\omega_0) =\frac{W^{-1}(\omega_0)_{j,k}R_k}{R_n W^{-1}(\omega_0)_{n,m}R_m},
\eeq
where 
\beq
W(\omega_0)_{j,k} = \int^{\infty}_0 d\omega (\omega-\omega_0)^2 K(\tau_j,\omega) K(\tau_k,\omega), \quad R_i = \int^{\infty}_0 d\omega K(\tau_i,\omega).
\eeq
The matrix $W$ typically has a condition number of $C(W) \equiv \lambda_{\text{max}}\\lambda_{\text{min}} \approx O(10^{20})$ and is thus extremely ill-conditioned. Previous studies have tried to remedy this situation by using a convariance matrix regularization \cite{Brandt:2015sxa,Boyda:2016emg}. 

In Tikhonov regularization, which we employ here, one casts the AC as a modified least-squares problem, $\text{min} \left( \Vert A x - b \Vert^2_2 + \Vert \Gamma x \Vert^2_2 \right)$, where $\Gamma$ is an appropriately chosen matrix which is here taken to be proportional to the identity $\lambda {\bm 1}$ \cite{Tikhonov}. Using the singular value decomposition (SVD) for the inverse matrix $W^{-1}$, the Tikhonov regularization results in the following modification
\beq
W^{-1}_{TR} = V D_{TR} U^{\top},~D_{TR} = \text{diag}( (\sigma_1+ \lambda)^{-1},\dots, (\sigma_N + \lambda)^{-1}) ),
\eeq
where $\sigma_i$ are the original singular values that are now smoothly cutoff for $\lambda \gg \sigma_i$. Comparing with other regularizations, we find that for a desired resolution in frequency space, Tikhonov regularization yields the spectral function with the smallest error. In \cite{Ulybyshev:2017ped}, this method was used to study the collective charge excitations of the extended Hubbard-Coulomb model. We refer the reader there for a more detailed analysis of the method.

\section{Results}\label{sec-2}
In this section we present our results for the Hubbard-Coulomb model on the square lattice and the Hubbard model on the hexagonal lattice. In the case of the Hubbard-Coulomb model, we were able to directly demonstrate the onset of the Mott-Hubbard transition by examining the momentum-averaged density of states. It is obtained from the following Euclidean correlator
\beq \label{PropagatorDOS}
G(\tau) = - \sum_x \vev{ T c(x,\tau) c^{\dagger}(x,0) }. 
\eeq
For weakly interacting fermions, Fermi-liquid theory predicts low energy excitations characterized by a dispersion relation with renormalized parameters, thus the well-defined Fermi surface still exists and there is a non-vanishing quasiparticle spectral weight, $Z$, at the Fermi energy which corresponds to $\omega = 0$ \cite{lifshitz2013statistical}. In Fig. (1a), we see how at small coupling we encounter such a situation. We have a large spectral weight at the origin which is a signature of the metallic phase. 

However, as the coupling increases, we encounter a situation where the quasiparticle spectral weight vanishes at the Fermi level and a new peak forms at finite $E$. This can be seen in Fig. (1b). 
\begin{figure}[thb] 
  \centering
  \includegraphics[width=14cm,clip]{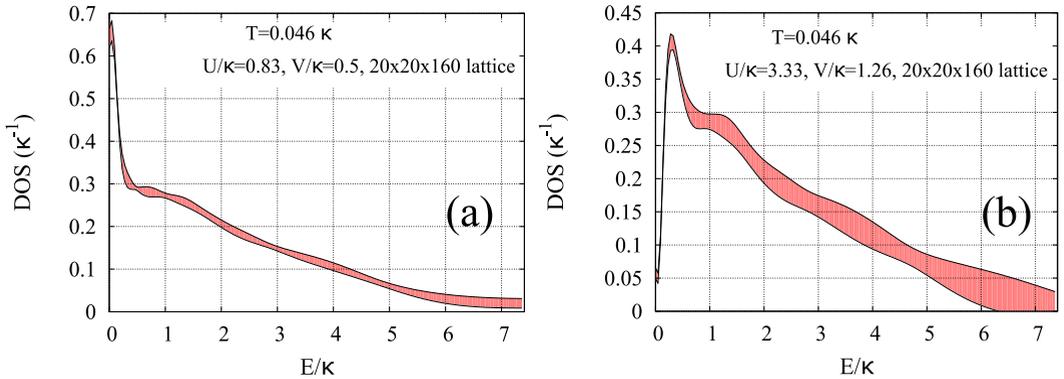}
  \caption{The momentum-averaged DOS in the Hubbard-Coulomb model for a lattice size of $N_s= 20$, $N_{\tau} = 160$ and temperature $T = 0.046 \kappa$, where $\kappa$ is the hopping amplitude between nearest neighbors. $U$ is the on-site Hubbard coupling and $V$ characterized the strength of the long-range Coulomb tail (it is equal to the interaction between nearest neighbors). (a) Weak-coupling case: One can clearly see the quasiparticle peak at zero frequency which characterizes the Fermi liquid phase. (b) Strong-coupling case: The spectral weight has been shifted to the Hubbard bands and the quasiparticles are gapped. }
  \label{fig-1}
\end{figure}
This phenomenon is known as the formation of Hubbard bands and is one signature of the Mott-Hubbard transition. In this phase, charge transfer is suppressed and the well-defined Fermi surface is completely destroyed. Being that this phenomenon occurs at strong coupling and is of a non-perturbative nature, lattice methods naturally lend themselves to the problem. The success of these methods can be compared to other methods commonly applied by condensed-matter theorists to the problem of the Mott-Hubbard transition such as dynamical mean-field theory (DMFT) \cite{Georges:1996zz,PhysRevLett.113.246407}. The advantage of the lattice formulation over these methods comes from the ability to natually take into account long-range interactions which can be important in the study of collective charge excitations \cite{Ulybyshev:2017ped}.

The case of the Hubbard model on the hexagonal lattice is a bit different as the picture of the phase structure is still unsettled. Although early studies suggested the existence of the spin-liquid phase which is predicted to exist in the region of the phase diagram between the semimetal and antiferromagnetic insulator (AFMI) phases \cite{PhysRevLett.95.036403,PhysRevB.76.035125,Meng2010}, more recent studies have found no evidence for this phase \cite{SciRep2012,Assaad:2013xua}. We have attempted to address the location, in terms of the Hubbard coupling $V_{00}$, of the onset of the gap for the fermionic quasiparticles relative to the formation of a magnetically ordered state. In doing so, one can hope to make a statement about the possible existence of the spin-liquid phase on the hexagonal lattice.

\begin{figure}[thb] 
  \centering
  \includegraphics[width=7cm,clip]{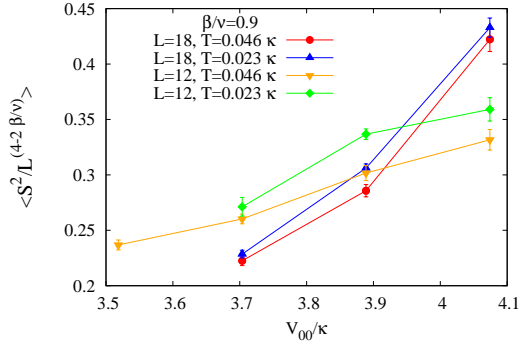}
  \caption{The magnetization on the hexagonal lattice is plotted as a function of on-site coupling for two lattice volumes, each at two different temperatures. A finite-scaling analysis for this correlator tells us that for a given temperature $T$, the coupling where the curves intersect defines the critical anti-ferromagnetic coupling $U_{\text{c,AF}}$.}
  \label{fig-3}
\end{figure}

To detect magnetic order, we have performed a finite scaling analysis of the long-range spin-spin correlations by measuring 
\beq \label{AFMOP}
\vev{ S^2 } \equiv \vev{  \sum_x \left(\hat{S}^z_x \right)^2 }, \quad \hat{S}^z_x \equiv \frac{1}{2} \left( \hat{c}^{\dagger}_{x,\uparrow} \hat{c}_{x,\uparrow} - \hat{c}^{\dagger}_{x,\downarrow} \hat{c}_{x,\downarrow} \right),
\eeq
where the sum runs over all spatial sites of the lattice and $\hat{S}^z_x$ refers to the third component of the spin operator at site $x$. This quantity should exhibit universal behavior at the phase transition point: $\vev{ S^2 }/L^{4-2\beta/\nu} \sim C$, where the constant $C$ does not depend on the lattice size $L$ and the coefficients $\beta$ and $\nu$ are the critical exponents for the corresponding phase transition. Thus, we can determine the critical coupling $U_{\text{c,AF}}$ by plotting the quantity $\vev{ S^2 }/L^{4-2\beta/\nu}$ for different lattice sizes and looking for the intersection point. The critical exponents for the AFM transition on the hexagonal lattice were found in \cite{Assaad:2013xua}.
In Fig. (\ref{fig-3}), one sees the results for (\ref{AFMOP}) for two spatial lattice sizes and two temperatures. From our results, one can see that the critical coupling lies somewhere between $3.9$ and $4$ (in units of the hopping parameter $\kappa$).

\begin{figure}[thb] 
  \centering
  \includegraphics[width=14.5 cm,clip]{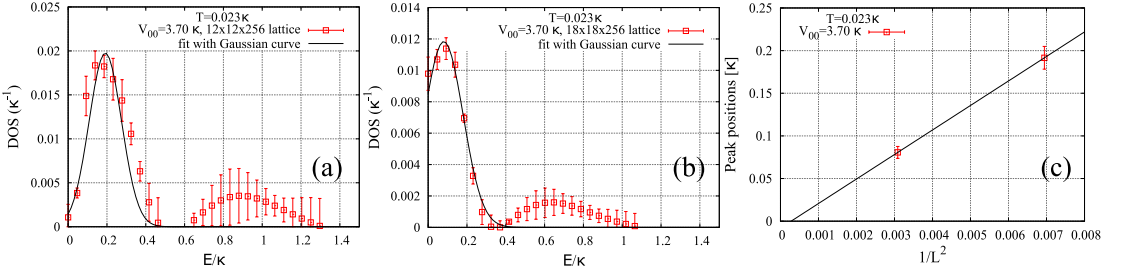}
  \caption{(a) The DOS at the $K$-point for $V_{00}=3.7\kappa$, $T=0.023 \kappa$, and $L=12$. (b) The DOS for the same parameters at spatial lattice size $L=18$. (c) The estimated peak positions from (a) and (b) plotted as a function of $1/L^2$.}
  \label{fig-4}
\end{figure}

\begin{figure}[thb] 
  \centering
  \includegraphics[width=14.5cm,clip]{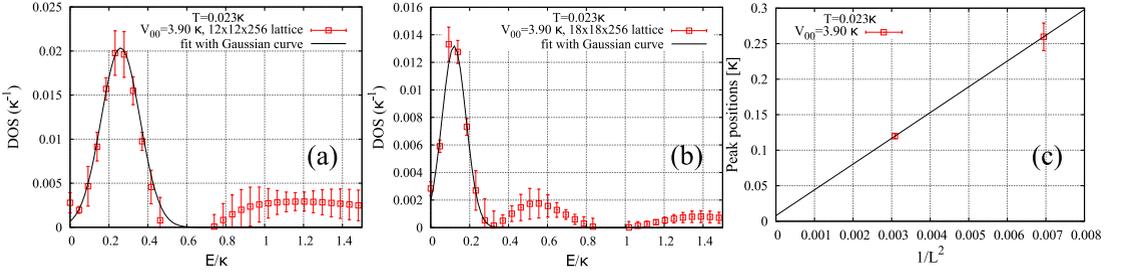}
  \caption{(a) The DOS at the $K$-point for $V_{00}=3.9\kappa$, $T=0.023\kappa$, and $L=12$. (b) The DOS for the same parameters at spatial lattice size $L=18$. (c) The estimated peak positions from (a) and (b) plotted as a function of $1/L^2$.}
  \label{fig-5}
\end{figure}

Now, to determine where the fermionic excitations become gapped, we look at the DOS. However, unlike the case of the square lattice, where at half-filling the Fermi surface was a diamond in the first Brillouin zone (BZ), for the hexagonal lattice at half-filling there are only two "Fermi points" where the valence and conduction bands touch. These are denoted by $K$ and $K'$ and are located at the corners of the BZ. It is well known from the theory of graphene that the low-energy excitations around these points are massless Dirac fermions. Thus, to determine if the excitations are gapped, we must look at the momentum-resolved DOS calculated exactly at the $K$-point. This was done at two different spatial volumes and we have plotted the position of the peak in $\omega$ as a function of the inverse lattice size $1/L^2$ for $V_{00}/\kappa=3.7, \,3.9$ in Figs. (\ref{fig-4}) and (\ref{fig-5}). Our results suggest that the gap may vanish in the thermodynamic limit although, admittedly, one would need several other lattice sizes to make a definitive conclusion. 

\section{Conclusion}\label{sec-3}
The accurate and systematic study of exotic phases of many-body systems and phase transitions that do not admit a local order parameter necessitates the use of innovative methods. For the two physical situations considered in this work, the gapped spin-liquid phase of the hexagonal lattice Hubbard model and the Mott-Hubbard transition on the square lattice, the ability to extract real-frequency information from Euclidean correlators through our modified BG method for AC was crucial. Namely, without making any assumptions about the spectral functions nor requiring unrealistically accurate Monte Carlo data, one is able to make quantitative predictions about the nature of the fermionic excitations. In particular, for the Hubbard model on the hexagonal lattice, our preliminary results suggest that the quasiparticles become gapped at a coupling that is comparable to that of the critical coupling at which system magnetically orders, $U_{\text{c,AF}}$, thus eliminating the gapped spin liquid phase. Future studies will attempt to get a more accurate determination of finite volume corrections in both the determination of the gap and $U_{\text{c,AF}}$.

\section*{Acknowledgements}
This work was partially supported by the HPC Center of Champagne-Ardenne ROMEO.
MU acknowledges support from DFG grant BU 2626/2-1. SZ acknowledges support by the National Science Foundation (USA) under grant PHY-1516509 and by the Jefferson Science Associates, LLC under U.S. DOE Contract $\#$ DE-AC05-06OR23177. CW is supported by the University of Kent, School of Physical Sciences.

\end{document}